\documentclass[prb,twocolumn,showpacs,amsmath,amssymb,superscriptaddress,bibnotes,floatfix]{revtex4-2}
\usepackage[dvipsnames]{xcolor}
\usepackage[T1]{fontenc}
\usepackage{amsmath}
\usepackage{braket}
\usepackage{simpler-wick}
\usepackage{tikz}
\usetikzlibrary{decorations.pathreplacing,decorations.markings, calc, shadows}
\tikzset{
  on each segment/.style={
    decorate,
    decoration={
      show path construction,
      moveto code={},
      lineto code={
        \path [#1]
        (\tikzinputsegmentfirst) -- (\tikzinputsegmentlast);
      },
      curveto code={
        \path [#1] (\tikzinputsegmentfirst)
        .. controls
        (\tikzinputsegmentsupporta) and (\tikzinputsegmentsupportb)
        ..
        (\tikzinputsegmentlast);
      },
      closepath code={
        \path [#1]
        (\tikzinputsegmentfirst) -- (\tikzinputsegmentlast);
      },
    },
  },
  mid arrow/.style={postaction={decorate,decoration={
        markings,
        mark=at position 0.53 with {\arrow[#1]{stealth}}
      }}},
}
\definecolor{viridisGreen}{RGB}{94, 201, 98}
\usepackage{graphicx}
\usepackage{hyperref}
\usepackage{placeins}
\usepackage{float}

\renewcommand{\d}{\mathrm{d}}
\newcommand{\ham}{\mathcal{H}}

\begin{document}
\preprint{APS/123-QED}

\title{Continuous Unitary Transformations using Tensor Network Representations Access the Full Many-
Body Localized Spectrum}

\author{Qiyu Liu}
 \email{qiyu.liu@tu-braunschweig.de}
 \affiliation{Technische Universität Braunschweig, Institut für Mathematische Physik,
 Mendelssohnstraße 3, 38106 Braunschweig, Germany}
 
\author{Jan-Niklas Herre}%
 \affiliation{Institute for Theory of Statistical Physics, RWTH Aachen University}
 
\author{Dante M.~Kennes}
  \email{dante.kennes@rwth-aachen.de}
 \affiliation{Institute for Theory of Statistical Physics, RWTH Aachen University}
 \affiliation{Max Planck Institute for the Structure and Dynamics of Matter,
Center for Free Electron Laser Science, Luruper Chaussee 149, 22761 Hamburg, Germany}

\author{Christoph Karrasch}
  \email{c.karrasch@tu-braunschweig.de}
 \affiliation{Technische Universität Braunschweig, Institut für Mathematische Physik,
 Mendelssohnstraße 3, 38106 Braunschweig, Germany}
\date{\today}

\begin{abstract}
We develop variational continuous unitary transformations (VCUTs), which integrate Wegner-Wilson
flow equations with tensor network techniques to approximately diagonalize many-body localized
(MBL) Hamiltonians. The diagonalizing unitary is represented as a matrix product operator whose
bond dimension controls the accuracy. For the disordered Heisenberg chain, VCUTs accurately
reproduces the full spectrum across the ergodic-to-MBL crossover at small system sizes and scales
to $L = 48$ sites. Beyond eigenenergies, the method can track the spatial entanglement structure of
the diagonalizing unitary $U(l)$ at each flow step, enabling identification
of local integrals of motion deep in the MBL phase.
\end{abstract}

\maketitle

\section{Introduction}
As one of the most intriguing phenomena in the field, many-body localization (MBL)
is intrinsically different from Anderson localization due to its emergent integrability, which
robustly prevents thermalization and conductivity even in the presence of interactions
\cite{Fleishman_1980,Basko_2006,Gornyi_2005,Nandkishore_2015,Abanin_2019,_nidari__2008,Oganesyan_2007,Pal_2010}. The fully many-body localized (fMBL) phase is characterized by a complete set of $L$ quasilocal
integrals of motion (LIOMs) $\{\tau_i^z\}_{i=1}^{L}$, known as $l$-bits
\cite{Ros_2015,Serbyn_2015}. These $l$-bits can be constructed from the physical spin
operators $\sigma_i^z$ via a quasilocal unitary transformation $U$ that diagonalizes the
Hamiltonian $\ham$: $\tau_i^z = U^\dagger \sigma_i^z U$. The operator $U$
connects the physical basis to the $l$-bit basis in which the Hamiltonian takes a diagonal form:
\begin{equation}
    \ham = \sum_i h_i \tau_i^z + \sum_{i<j} J_{ij} \tau_i^z \tau_j^z + \cdots\,,
    \label{eq:lbit_hamiltonian}
\end{equation}
where the couplings $J_{ij}$ decay exponentially with distance $|i-j|$. As evident from this form,
the $l$-bits are mutually commuting operators that also commute with $\ham$, giving rise to the
emergent integrability of the fMBL phase. Crucially, both $U$ and the $l$-bits exhibit spatial
locality: the $l$-bits are exponentially localized around their associated lattice sites, and $U$
can be approximated by a finite-depth quantum circuit and admits an efficient MPO representation
\cite{Pollmann2016,Wahl2017,Pekker_2017}. Consequently, applying $U$ to the $2^L$ product states
in the $l$-bit basis yields all eigenstates with area-law entanglement. Furthermore, the exponentially decaying interactions $J_{ij}$ between $l$-bits lead to slow
dephasing dynamics, which governs the spreading of local operators. In fMBL systems, a local
operator spreads only logarithmically in time, $r \sim \xi \log(t)$, where $\xi$ is the
localization length \cite{Luitz_2015,Luitz_2017,Altman_2015}. In contrast, ergodic systems exhibit ballistic operator spreading, bounded by the Lieb-Robinson velocity \cite{Lieb_1972}, which sets the maximum speed at which information can propagate in systems with local interactions.

Verifying the existence of $l$-bits and understanding the ergodic-to-MBL transition requires probing
the properties of highly excited eigenstates. This can be achieved either by computing eigenstates
directly or by investigating the time evolution of many-body systems, which reveals whether a system
thermalizes or not. Both approaches face significant challenges due to the exponential growth of the Hilbert space, 
severely limiting accessible time scales and system sizes.

Exact diagonalization (ED) provides numerically exact results but is limited to small system sizes
($L \lesssim 20$). The shift-and-invert technique \cite{Pietracaprina_2018} extends ED to larger systems
($L \lesssim 26$) by targeting eigenstates at specific energies. However, this method requires
inverting the shifted Hamiltonian, typically performed via LU decomposition. Since the sparsity
pattern of the Hamiltonian is not preserved during decomposition, the resulting memory overhead
becomes the main bottleneck, and the method remains restricted to computing individual eigenstates
rather than the full spectrum.

Tensor network methods offer an alternative route. DMRG-X \cite{Khemani_2016} adapts the density
matrix renormalization group to target highly excited eigenstates by exploiting their low
entanglement in the MBL phase. However, the method requires a good initial guess and converges
reliably only deep in the localized regime where eigenstates resemble product states. Near the
transition, increased entanglement limits its applicability. Unitary tensor network approaches
\cite{Wahl2017, Pollmann2016, Pekker_2017} represent the diagonalizing transformation $U$ as a
tensor network, providing access to all eigenstates simultaneously. These methods exploit the
finite-depth circuit structure of $U$ in the MBL phase but rely on local optimization schemes that
can become trapped in local minima, and their accuracy degrades as the system approaches the
ergodic-to-MBL transition.

Continuous unitary transformations (CUTs) \cite{Wegner_2000,Kehrein_2006} offer an alternative by
letting the Hamiltonian flow toward a diagonal form via coupled differential equations. However, this
system of equations is not closed, and the number of terms proliferates during the flow,
necessitating truncation schemes \cite{Fischer_2010, Thomson_2018}. The recently
developed Tensor Flow Equations (TFE) \cite{Thomson_2024, Herre2024} use an
expansion in interaction strength, yielding promising results for weakly interacting systems but
becoming less accurate at strong correlations. Previous attempts to combine tensor networks with CUTs include time-evolving block decimation
(TEBD)-based approaches
\cite{Sahin2017} and Baker-Campbell-Hausdorff (BCH)-based unitary
tensor networks
\cite{yu_2019}. However, obtaining the complete spectrum with controlled accuracy using
tensor-network-based flow equations has remained out of reach.

In this work, we present variational continuous unitary transformations (VCUTs), a method that
combines tensor network representations with Wegner-Wilson flow equations using the time-dependent
variational principle (TDVP) algorithm \cite{Haegeman2011, Haegeman2016}. Unlike TEBD, TDVP
naturally handles the long-range interactions generated during the flow. By representing the
diagonalizing unitary as a matrix product operator (MPO) with controlled bond dimension, VCUTs
provides an optimized truncation scheme based on entanglement, enabling access to the entire spectrum with controlled accuracy.

We benchmark VCUTs against ED and TFE for system sizes $L \leq 16$, demonstrating accurate
approximation of the full spectrum across the ergodic-to-localized crossover. The method maintains accuracy for system sizes up to
$L=48$, well beyond the reach of ED. Furthermore, VCUTs provides direct access to the dynamics of the
diagonalizing unitary transformation and the spreading of local operators, offering real-space
resolved insights into the localization transition. We observe that spatially localized regions of
high entanglement persist deep into the localized regime, concentrated at bonds
with small local disorder gradient $\Delta h_i$, directly reflecting the resonance structure of the
MBL phase, and reliably identify LIOMs in this regime.

The paper is organized as follows: Section~\ref{section: model} introduces the model, and
Section~\ref{section: Method} describes the VCUTs method. Results are presented in
Section~\ref{section: Results}, followed by a discussion in Section~\ref{section: Discussion}.

\section{The Model}\label{section: model}

We study the disordered spin-$\frac{1}{2}$ XXZ chain with open boundary conditions, a paradigmatic
model for investigating many-body localization. The Hamiltonian reads
\begin{equation}
    \ham = \sum_{j=1}^{L-1} J_{0}\left(S^{x}_{j}S^{x}_{j+1}+S^{y}_{j}S^{y}_{j+1}+\Delta
    S^{z}_{j}S^{z}_{j+1}\right) + \sum_{j=1}^{L} h_{j}S^{z}_{j}\,,
    \label{eq:spinful_XXZmodel}
\end{equation}
where $J_0$ is the exchange coupling, $\Delta$ controls the interaction anisotropy, and the
on-site fields $h_j$ are drawn independently from a uniform distribution $h_j \in [-W, W]$ with
disorder strength $W$. Throughout this work, we set $J_0 = 1$ and $\Delta = 0.5$.

The isotropic case $\Delta = 1$ corresponds to the random-field Heisenberg chain (RFHC), a canonical model for MBL with the ergodic-to-MBL transition previously estimated at critical disorder strength $W_c \approx 3.7$~\cite{Luitz_2015,Pal_2010,ifmmodeSelseSfiuntajs2020,ifmmodeSelseSfiuntajs2020a,Colbois2024}. The critical disorder depends on the interaction strength
$\Delta$~\cite{Hopjan2024}: weaker interactions bring the system closer to the non-interacting
Anderson localized limit, such that for $\Delta = 0.5$ the transition is expected at a lower value
of $W_c$. We note, however, that finite-size effects in numerical studies remain a concern. The
observation of slow dynamics toward delocalization deep in the putative MBL phase has raised
questions about the stability of the MBL phase in the thermodynamic
limit~\cite{Weiner2019,Doggen2018,Zisling_2022}.
\section{The Method} \label{section: Method}

\subsection{Continuous unitary transformations}\label{section: CUT}

Continuous unitary transformations (CUTs) provide a framework for diagonalizing quantum
Hamiltonians by flowing them toward diagonal form~\cite{Wegner_2000, Kehrein_2006}. The flow is
governed by the differential equation
\begin{equation}
\label{eq:WWF}
    \frac{\d \ham(l)}{\d l} = \left[\eta(l), \ham(l)\right],
\end{equation}
where $l$ is the flow parameter and $\eta(l)$ is an anti-Hermitian generator. The formal solution
is given by
\begin{equation}
    \ham(l) = U(l)\ham(0)U^{\dagger}(l)\,,
    \label{eq:unitary transformation}
\end{equation}
with $U(l) = \mathcal{T}_l \exp\left(- \int_{0}^{l} \eta(l') \d l'\right)$, where $\mathcal{T}_l$
denotes flow-ordering. A common choice is the Wegner generator~\cite{Wegner_2000},
\begin{equation}
\label{eq:Wegner generator}
    \eta(l) = \left[\ham_\text{d}(l), \ham_{\text{od}}(l)\right],
\end{equation}
where $\ham_\text{d}(l)$ and $\ham_{\text{od}}(l)$ denote the diagonal and off-diagonal parts of
the Hamiltonian, respectively, as illustrated in Fig.~\ref{fig:demo for the method}(b). This
choice guarantees that the off-diagonal norm decreases monotonically, driving the Hamiltonian
toward diagonal form as $l \to \infty$. Once the flow converges, the eigenvalues can be read off
from the diagonal elements of $\ham(l)$, and the diagonalizing unitary $U(l)$ connects eigenstates
to the computational basis.

\subsection{Tensor network formulation of the flow}\label{section: algorithm}

\begin{figure}[t]
    \centering
    \includegraphics[width=1.\columnwidth]{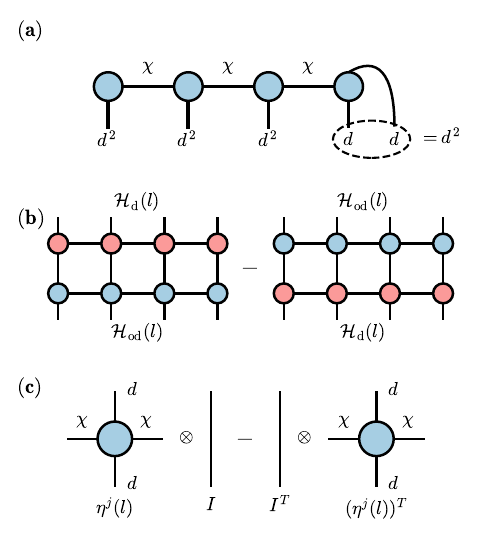}
\caption{\label{fig:demo for the method}
Tensor network representation of the VCUTs method. (a)~Vectorization of an MPO into an MPS by
combining the bra and ket physical indices at each site into a single index of dimension $d^2$.
(b)~Construction of the Wegner generator $\eta(l) = [\ham_\text{d}(l), \ham_{\text{od}}(l)]$ as
the commutator of the diagonal and off-diagonal parts of the Hamiltonian. (c)~The local
superoperator $\mathcal{U}^{j}(l)$ from Eq.~\eqref{eq:superoperator}.}
\end{figure}

The key idea of VCUTs is to represent the flowing Hamiltonian $\ham(l)$ as a matrix product
operator (MPO) and solve the flow equation~\eqref{eq:WWF} using tensor network techniques. At each
infinitesimal flow step, the Hamiltonian evolves according to
\begin{equation}
\label{eq:flow step}
\ham(l+\d l) = e^{\eta(l)\d l} \ham(l) e^{-\eta(l)\d l}.
\end{equation}
To cast this into a form amenable to standard MPS time-evolution algorithms, we vectorize the MPO
by combining the bra and ket indices at each site into a single index of dimension $d^2$, yielding
a matrix product state (MPS) representation of the vectorized Hamiltonian. This vectorization
technique is analogous to that used for matrix product density operators
(MPDOs)~\cite{Zwolak_2004, Werner_2016, Guth_2020, Weimer_2021}. Formally, we map
$\ham(l) \mapsto \vert \ham(l)\rangle\rangle$, as illustrated in Fig.~\ref{fig:demo for the
method}(a). This vectorization transforms Eq.~\eqref{eq:flow step} into a Schrödinger-like
equation,
\begin{equation}
\label{eq:Schrodinger-like}
\frac{\d \vert \ham(l)\rangle\rangle}{\d l} = \mathcal{U}(l)\vert \ham(l)\rangle\rangle,
\end{equation}
where $\mathcal{U}(l)$ is a superoperator that acts on the vectorized Hamiltonian. The local
tensor components of $\mathcal{U}(l)$ on site $j$ are given by
\begin{equation}
\mathcal{U}^{j}(l) = \eta^{j}(l) \otimes I - I \otimes (\eta^{j}(l))^{T},
\label{eq:superoperator}
\end{equation}
where $\eta^{j}(l)$ is the local tensor of the Wegner generator defined in
Eq.~\eqref{eq:Wegner generator}, as shown graphically in Fig.~\ref{fig:demo for the method}(c).
The superoperator $\mathcal{U}(l)$
can itself be represented as an MPO acting on the enlarged local Hilbert space of dimension $d^2$;
we refer to this as a super-MPO (SMPO)~\cite{Zwolak_2004, Werner_2016, Guth_2020, Weimer_2021}.

To solve Eq.~\eqref{eq:Schrodinger-like}, we employ the time-dependent variational principle (TDVP)
algorithm, which provides a stable and efficient framework for time-evolving
MPS~\cite{Haegeman2011, Haegeman2016}.Since the
superoperator $\mathcal{U}(l)$ is generally non-Hermitian, we employ the Krylov-Arnoldi algorithm
to compute the required matrix exponentials~\cite{Hochbruck1997, Saad1992}.

\subsection{Diagonalizing unitary and local integrals of motion}\label{section: compute Op flow}

Beyond the eigenvalue spectrum, VCUTs provides direct access to the diagonalizing unitary $U(l)$
and the local integrals of motion (LIOMs).

\textit{a. Diagonalizing unitary.}---To construct the diagonalizing unitary $U(l)$ within the
tensor network framework, we start from the MPO representation of the identity operator
\begin{equation}
I = \sum_{\sigma} \vert \sigma \rangle \langle \sigma \vert,
\label{eq:identity MPO}
\end{equation}
where the sum runs over all computational basis states $\sigma = (\sigma_1, \ldots, \sigma_L)$.
The identity MPO has bond dimension 1 with local tensors $\delta_{\sigma_j \sigma'_j}$.
We vectorize this MPO into an MPS representation $\vert I \rangle\rangle$ and evolve it
alongside the Hamiltonian using the same TDVP setup. At each infinitesimal step, $U(l)$ updates
via left multiplication by $e^{\eta(l)\d l}$. In the vectorized representation, this is
implemented using a modified superoperator with local components
\begin{equation}
\mathcal{U}_{U}^{j}(l) = \eta^{j}(l) \otimes I,
\label{eq:unitary superoperator}
\end{equation}
yielding the vectorized diagonalizing unitary $\vert U(l) \rangle\rangle$.

\textit{b. Local integrals of motion.}---In the MBL phase, the LIOMs $\tau_j^z$ are quasilocal
operators that commute with each other and with the Hamiltonian~\cite{Serbyn_2013,
Ros_2015}. They are related to the bare spin operators by the diagonalizing unitary,
\begin{equation}
\tau_j^z = U^\dagger \sigma_j^z U.
\label{eq:LIOM}
\end{equation}
Once the flow converges and the final diagonalizing unitary $U$ is obtained, we compute the LIOMs
by applying $U$ to the MPO representation of the bare spin operator $\sigma_j^z$.

\subsection{Implementation details}\label{section: implementation}

In our implementation, we represent the vectorized Hamiltonian with a maximum bond dimension
$D_H$. The Wegner generator $\eta(l)$ is represented as an MPO with bond dimension $D_\eta = D_H$,
as illustrated in Fig.~\ref{fig:demo for the method}(b). The diagonalizing unitary $U(l)$ requires
a slightly larger bond dimension to maintain accuracy throughout the flow, and we use
$D_U = D_H + 8$. To assess convergence with respect to bond dimension, we test
$D_H = 16, 32, 48$ in our benchmarks.

The progress of the flow is monitored via the off-diagonal variance
\begin{equation}
   V(l) = \frac{1}{2^{L}}\sum_{i \neq j}|\ham_{ij}(l)|^{2},
   \label{eq:variance}
\end{equation}
which can be computed efficiently using tensor network contractions. The flow is terminated when
the variance decreases below a threshold $V(l) < 10^{-5} V(0)$ relative to the initial variance,
or when the variance decrease stalls (less than $0.05\%$ reduction over 20 consecutive steps).

We employ an adaptive time-stepping scheme that dynamically adjusts the flow step $\d l$ based on
the variance reduction per step. At each step, the algorithm tests three candidate step sizes
($0.9\,\d l$, $\d l$, $1.1\,\d l$) in parallel and selects the one yielding optimal variance
reduction. The target is a variance drop of $1$--$3\%$ per step: if the drop exceeds $3\%$, the
step size is reduced by a factor of $0.8$; if below $1\%$, it is increased by a factor of $1.2$.

The TDVP algorithm introduces several sources of error~\cite{Haegeman2016}: (i)
projection error arising from the finite bond dimension of the MPS ansatz, (ii) time-step error of
order $\mathcal{O}(\d l^3)$ per step from the second-order integrator, (iii) truncation error from
bond dimension reduction in TDVP2, and (iv) local solver error from the numerical computation of
matrix exponentials. For the local solver, we use the Krylov-Arnoldi algorithm with tolerance
$10^{-12}$, Krylov dimension $20$--$40$ (scaled with system size) rendering the solver error negligible.  
We adopt a two-stage approach: initially, we use two-site
TDVP to allow the bond dimension to grow dynamically as the flow progresses. Once the
bond dimension reaches the specified cutoff $D_H$, we switch to one-site TDVP for the
remainder of the flow. This hybrid strategy balances flexibility in capturing the growing
entanglement with the computational efficiency and stability of single-site updates. The dominant
sources of error are therefore {projection error} and time-step error , which we can control through careful 
selection of $D_H$ and adaptive time-stepping.

For computing the autocorrelation function, the transformed states $\ket{\psi_n} = U\ket{n}$
obtained from VCUTs are generally non-orthogonal. To extract proper eigenstates, we solve the
generalized eigenvalue problem $H_{mn} c_k^{(n)} = E_k S_{mn} c_k^{(n)}$, where $H_{mn} = \langle
\psi_m | \ham | \psi_n \rangle$ and $S_{mn} = \langle \psi_m | \psi_n \rangle$ are the Hamiltonian
and overlap matrices. We orthogonalize using Cholesky decomposition of the overlap matrix $S = LL^T$,
which is equivalent to Gram-Schmidt orthogonalization, transforming to $\tilde{H} = L^{-1} H
(L^{-1})^T$ and solving the standard eigenvalue problem.

\section{Results}\label{section: Results}

\begin{figure*}[t]
    \centering
    \includegraphics[width=0.95\textwidth]{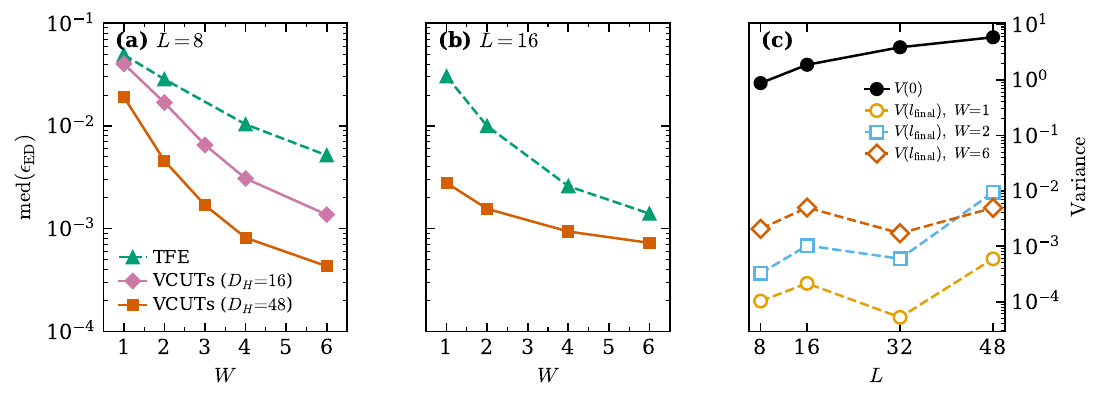}
\caption{\label{fig:median errors}
Mean of the median relative energy error with respect to ED as a function of
disorder strength $W$ [panels (a) and (b)], and variance suppression as a function of system size
$L$ [panel (c)].
(a)~$L=8$: TFE (triangles), VCUTs with $D_H = 16$ (diamonds) and $D_H = 48$ (squares), averaged
over 50 disorder realizations.
(b)~$L=16$: TFE (triangles) and VCUTs with $D_H = 48$ (squares), averaged over 30 disorder realizations.
(c)~Initial variance $V(0)$ (black circles, solid line) and final variance $V(l_{\mathrm{final}})$
(open markers, dashed lines) as a function of system size $L$ for disorder strengths $W=1, 2, 6$.
For $L=8$, $D_H = 32$, averaged over 49 disorder realizations;
for $L=16$, $D_H = 48$, averaged over 30 realizations;
for $L=32$ and $48$, $D_H = 32$, a single realization is shown.}
\end{figure*}

We benchmark VCUTs on the disordered XXZ chain defined in Eq.~\eqref{eq:spinful_XXZmodel} with
$\Delta = 0.5$. The implementation parameters are specified in
Section~\ref{section: implementation}.

\subsection{Full spectrum in the localized regime}

\begin{figure}[tb]
    \centering
    \includegraphics[width=1.0\columnwidth]{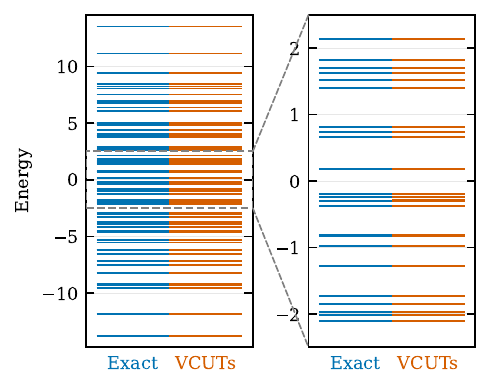}
\caption{\label{fig:full spectrum comparison}
Comparison between exact eigenenergies (blue lines) and energies obtained via VCUTs (red lines) for
a system with $L=8$ and $W=6$. The left panel shows all 70 eigenstates in the zero magnetization
sector. The right panel displays a magnified view of the energy levels near zero.}
\end{figure}

We first compare the eigenenergies obtained by VCUTs with exact diagonalization (ED). Within VCUTs,
the eigenenergies are read directly from the diagonal elements of the flowed Hamiltonian. In
Fig.~\ref{fig:full spectrum comparison}, we show results for a single disorder realization with
$L=8$ and $W=6$. The left panel displays all 70 eigenstates in the zero magnetization sector, while
the right panel provides a magnified view of the energy levels near zero. The VCUTs results agree
well with ED throughout the spectrum.

Next, we quantitatively benchmark VCUTs against ED and Tensor Flow Equations (TFE), another
CUTs-based method capable of accessing the full spectrum. TFE has been successfully applied to
finite-size quantum many-body systems, in some cases reaching sizes beyond
ED~\cite{Thomson2023a, Thomson_2024, Herre2024}. For comparison, we use TFE with
the exact same implementation as in Ref.~\cite{Herre2024}; for details see Appendix~\ref{sec: TFE}. To quantify the accuracy, we
define the relative energy error for each eigenstate $\ket{j}$ as
\begin{equation}
   \epsilon^{(j)}_\text{ED} = \left|\frac{E^{(j)}_\text{CUTs} - E^{(j)}_\text{ED}}{E^{(j)}_\text{ED}} \right| \,,
   \label{eq: TFE_ED energy error}
\end{equation}
and compute the median error over all eigenstates in the zero magnetization sector,
$$\mathrm{med}(\epsilon_\text{ED}) \equiv \underset{j}{\mathrm{med}}(\epsilon^{(j)}_\text{ED})\;.$$

In Fig.~\ref{fig:median errors}, we compare the median errors for TFE and VCUTs across a range of
disorder strengths. For $L=8$ [panel (a)], VCUTs with bond dimension $D_H = 16$ already outperforms
TFE for all values of $W$. Increasing the bond dimension to $D_H = 48$ further improves the
accuracy, with the most significant gains observed in the localized regime at large $W$. The same
comparison at $L=16$ [panel (b)] confirms that VCUTs maintains its advantage over TFE at larger
system sizes, with the error consistently a factor of $2$--$3$ smaller across all disorder strengths.
This is despite TFE employing a state-of-the-art generator optimized for the small-disorder
regime~\cite{Herre2024}, whereas VCUTs uses only the standard Wegner generator.
To assess the scalability of VCUTs, we examine the variance suppression across
different system sizes. In Fig.~\ref{fig:median errors}(c), we show both the initial variance
$V(0)$ and the final variance $V(l_{\mathrm{final}})$ as a function of $L$ for several disorder
strengths. The initial variance grows linearly with system size, $V(0) \propto L$, reflecting the
extensive number of off-diagonal terms in the Hamiltonian. The final variance remains orders of
magnitude smaller than $V(0)$ across all system sizes and disorder strengths, demonstrating that
the Hamiltonian is approximately diagonalized even for $L=48$, well beyond the reach of ED.
\subsection{Entanglement structure of the flow}\label{section:results of Op flow}

A key feature of MBL systems is that the diagonalizing unitary $U$ can be represented as a
finite-depth quantum circuit~\cite{Pekker2017TN, Pekker2017}, implying an efficient MPO representation with bounded bond
dimension. In VCUTs, the full unitary is constructed as
a product of infinitesimal transformations,
\begin{equation}
U(l) = \prod_{l'=0}^{l} e^{\eta(l') \d l'} \equiv \prod_{l'=0}^{l} \d U(l'),
\label{eq:U_product}
\end{equation}
where each $\d U(l') = e^{\eta(l') \d l'}$ is an MPO. To compute $\d U(l')$, we apply the
superoperator from Eq.~\eqref{eq:unitary superoperator} with the generator $\eta(l')$ and step size
$\d l'$ to the identity operator, using TDVP2 with a truncation threshold of $10^{-4}$. The
entanglement structure of these infinitesimal generators provides insight into the spatial
organization of the flow.

\begin{figure*}[t]
    \centering
    \includegraphics[width=0.95\textwidth]{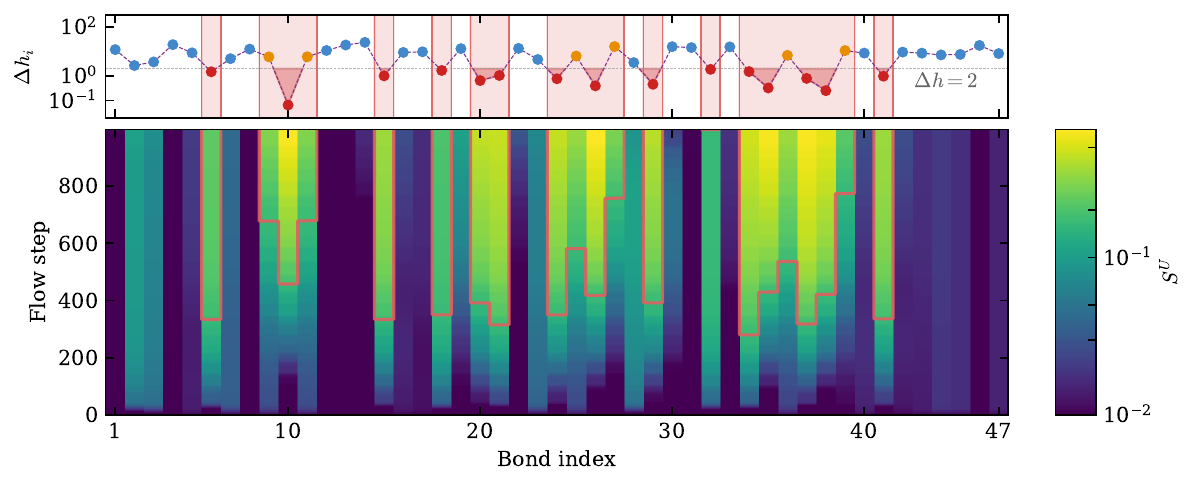}
\caption{\label{fig:MERA stucture of dU}
Spatially resolved bond entropy $S^U$ of the diagonalizing unitary $U(l)$ during the VCUTs flow
for $L=48$, $W=12$ (single disorder realization, $D_H=48$). \emph{Main panel:} $S^U$ on a
logarithmic color scale as a function of bond index (horizontal) and flow step (vertical). The red
contour encloses the high-entropy core ($S^U \geq S^U_{\min}$, where the threshold is set
adaptively from the minimum final $S^U$ among bonds with $\Delta h_i < 2$). Blue shading marks
bonds where $\Delta h_i < 2$. \emph{Top panel:} local disorder gradient
$\Delta h_i = |h_i - h_{i+1}|$ at each bond; scatter points are colored red (below $\Delta h$
threshold), orange (above threshold but inside the entropy core), or blue (above threshold and
outside the core). The entropy accumulates preferentially at low-$\Delta h$ bonds, revealing the
resonance-driven spatial structure of the flow.}
\end{figure*}

In Fig.~\ref{fig:MERA stucture of dU}, we show the bond entropy $S^U$ of the
diagonalizing unitary $U(l)$ as a function of bond index and flow step for $L=48$ and $W=12$, deep
in the MBL phase. The entropy $S^U$ grows predominantly at bonds where the local disorder gradient
$\Delta h_i = |h_i - h_{i+1}|$ is small (blue-shaded regions), corresponding to near-resonant
pairs of neighboring sites. These bonds are the first to accumulate entanglement as the flow
progresses, forming a high-entropy core (enclosed by the red contour). To delineate
this core we use the finite-size estimate $h_c \approx 2$ for the critical disorder gradient as
a resonance criterion: bonds with $\Delta h_i < h_c$ are classified as near-resonant (red points
in the top panel). We then set the entropy threshold $S^U_{\min}$ to the lowest bond entropy among
these near-resonant bonds at the final flow step, and define the ``bubble'' as the connected region
where $S^U \geq S^U_{\min}$.
In the top panel, bonds with $\Delta h_i > h_c$ that nonetheless lie inside the bubble are marked
in orange: although initially off-resonant, their bond entropy exceeds the threshold, indicating
that they have been ``heated'' by the adjacent high-entropy resonant (red) bonds.
Bonds with $\Delta h_i > h_c$ outside the bubble (blue points) remain largely unaffected. This spatial selectivity confirms that the
entanglement generated during the VCUT flow is governed by the local resonance structure of the
disorder landscape, providing a direct real-space picture of where the emergent integrals of motion
are localized.
\section{Real Time Dynamics}\label{section: real time dynamics}

Having established that VCUTs accurately reproduces the eigenenergies, we now investigate its
ability to capture real-time dynamics through the autocorrelation function. The dynamical
autocorrelation of a local operator provides a direct probe of the localization properties of the
system and serves as a stringent test of the quality of the LIOMs obtained from the diagonalizing
unitary.

We compute the infinite-temperature autocorrelation function of the $\sigma^z$ operator located at
the middle site $j = L/2$:
\begin{equation}
C(t) = \frac{1}{2^L} \mathrm{Tr}\left[\sigma_j^z(t) \sigma_j^z(0)\right],
\label{eq:autocorrelation}
\end{equation}
where $\sigma_j^z(t) = e^{i\ham t} \sigma_j^z e^{-i\ham t}$. Using the LIOM representation
$\tau_j^z = U^\dagger \sigma_j^z U$ obtained from VCUTs, the autocorrelation can be computed
efficiently. For comparison, we also compute the autocorrelation using ED (which provides the exact
result) and TFE.

\begin{figure}[t]
    \centering
    \includegraphics[width=0.95\columnwidth]{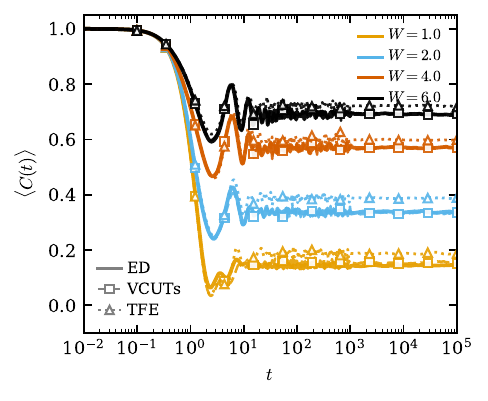}
\caption{Disorder-averaged dynamical autocorrelation $\langle C(t) \rangle$ for $L=8$ and three
disorder strengths $W=1, 2, 6$, averaged over 50 realizations. Solid lines: ED (exact). Dashed lines
with square markers: VCUTs ($D_H = 48$). Dotted lines with triangle markers: TFE. Different colors
correspond to different disorder strengths. A moving average is applied for $t > 1000$ to reduce
oscillations. The autocorrelation is computed from the LIOM $\tau^z_{L/2} = U^\dagger \sigma^z_{L/2}
U$ located at the middle site.}
\label{fig:autocorrelation}
\end{figure}

In Fig.~\ref{fig:autocorrelation}, we show the disorder-averaged autocorrelation for $L=8$ at three
representative disorder strengths spanning the ergodic-to-MBL crossover. At early times
($t \lesssim 10$), both VCUTs and TFE show excellent agreement with the exact ED results across all
disorder strengths. This demonstrates that the LIOMs obtained from both methods accurately capture
the short-time dynamics, which is governed by the local structure of the operators.

The long-time behavior reveals an interesting dependence on disorder strength. In the ergodic regime
($W=1$), VCUTs maintains good agreement with ED even at longer times, demonstrating that the method
captures the essential physics of the delocalized phase. At the crossover ($W=2$) and in the
localized regime ($W=6$), VCUTs and TFE exhibit similar long-time behavior, with the
autocorrelation approaching a finite saturation value characteristic of localized systems.

However, subtle deviations between VCUTs/TFE and ED emerge at very long times in the crossover and
localized regimes. This mismatch arises because the transformed states $\ket{\psi_n} = U\ket{n}$
obtained from VCUTs are not exactly orthogonal due to the finite bond dimension truncation. When
computing the autocorrelation, these non-orthogonality effects accumulate over many eigenstates and
manifest as deviations at long times. The Gram-Schmidt orthogonalization procedure described in
Section~\ref{section: implementation} mitigates but does not fully eliminate this effect, as it
relies on the approximate overlap matrix computed from the truncated MPS. Nevertheless, VCUTs
exhibits slightly better overall correspondence with ED compared to TFE, consistent with its
non-perturbative treatment of interactions.

\section{Insight on the ergodic-to-localization crossover}

A key signature of the ergodic-to-MBL transition is the behavior of eigenstate entanglement
entropy~\cite{Kjall2014, Luitz_2015, Serbyn2015EE}. In the ergodic phase, eigenstates exhibit
volume-law entanglement consistent with the eigenstate thermalization hypothesis (ETH), while in
the MBL phase, eigenstates follow an area law due to the localized nature of the LIOMs. Near the
transition, the sample-to-sample fluctuation of the entanglement entropy develops a
broad maximum, which has been proposed as a practical indicator of the crossover region and the
critical disorder strength $W_c$~\cite{Khemani_PRX_2017_CriticalProperties,Wahl2017}.

\begin{figure}[H]
    \centering
    \includegraphics[width=1.0\columnwidth]{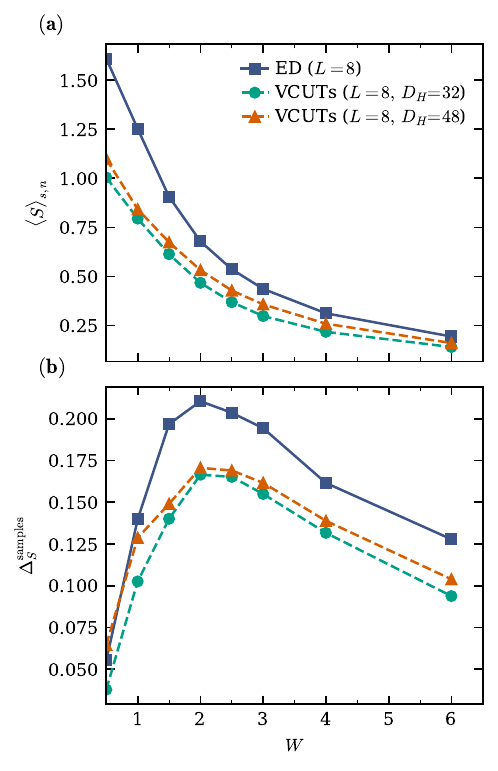}
\caption{Half-cut entanglement entropy statistics as a function of disorder strength $W$; for each
$W$ we average over $50$--$100$ disorder realizations.
(a)~Mean entanglement entropy $\langle S \rangle_{s,n}$ (defined in the text) from ED ($L=8$) and VCUTs ($L=8,\,D_H=32,48$).
(b)~Sample-to-sample fluctuation diagnostic $\Delta_S^{\mathrm{samples}}$ (defined in the text) for
ED ($L=8$) and VCUTs ($L=8,\,D_H=32,48$).}
\label{fig:halfcut_EE}
\end{figure}

In Fig.~\ref{fig:halfcut_EE}, we compare the half-cut entanglement entropy statistics obtained from
VCUTs with exact diagonalization for $L=8$. The VCUT dressed states are generated by applying the
diagonalizing unitary $U$ from the flow to product states $\ket{n}$, $\ket{\psi_n}=U\ket{n}$.
In panel (a), we plot the mean entanglement entropy
$\langle S \rangle_{s,n} = \langle\langle S_{cs}^{n}\rangle_{c,s}\rangle_n$,
where the half-chain entanglement entropy $S_{cs}^{n}$ is first averaged over eigenstates $s$
and spatial cuts $c$ for each disorder realization $n$, and then averaged over realizations.
In our case $c$ runs over the three most central left--right bond cuts (the middle bond and its
two nearest neighbors), which provides a robust estimate of the half-chain entanglement while
remaining accessible from the MPS bond structure.
$\langle S \rangle_{s,n}$ decreases monotonically with increasing disorder strength, reflecting
the crossover from volume-law to area-law entanglement.
In panel (b), we show the sample-to-sample fluctuation diagnostic
$\Delta_S^{\mathrm{samples}} = \sigma_n(\langle S_{cs}^{n}\rangle_{c,s})$,
following Refs.~\cite{Khemani_PRX_2017_CriticalProperties,Wahl2017}, where $\sigma_n$ denotes
the standard deviation across disorder realizations.
$\Delta_S^{\mathrm{samples}}$ thus quantifies how strongly the mean entanglement entropy varies
from sample to sample at fixed disorder strength; a pronounced peak signals enhanced heterogeneity
in the crossover regime.

$\Delta_S^{\mathrm{samples}}$ exhibits a broad maximum at intermediate disorder, as also observed in
Refs.~\cite{Khemani_PRX_2017_CriticalProperties,Wahl2017}. For
$L=8$, ED and VCUTs both display a qualitatively similar peak in panel (b). The remaining
deviations are most visible at low disorder, where entanglement is largest and finite bond dimension
effects are strongest.

Importantly, increasing the bond dimension from $D_H = 32$ to $D_H = 48$ systematically improves the
agreement with ED, particularly in the low-disorder regime where entanglement is largest. With
$D_H = 48$, VCUTs captures the qualitative features of the crossover, including the peak position
and the overall shape of the $\Delta_S^{\mathrm{samples}}$ curve. This demonstrates that VCUTs can
provide meaningful insights into the ergodic-to-MBL crossover, with accuracy that improves as the
bond dimension is increased.

\section{Discussion}\label{section: Discussion}

We have developed variational continuous unitary transformations (VCUTs), a method that combines
Wegner-Wilson flow equations with tensor network techniques to approximately diagonalize many-body
Hamiltonians. By representing the flowing Hamiltonian as an MPO and employing TDVP for the
integration, VCUTs provides a controlled truncation scheme based on entanglement that prevents the
proliferation of terms inherent to conventional CUTs implementations. The bond dimension $D_H$
serves as the single control parameter that governs the trade-off between accuracy and computational
cost.

We benchmarked VCUTs on the disordered XXZ chain across the ergodic-to-MBL crossover. For small
system sizes ($L=8$), VCUTs accurately reproduces the full spectrum, outperforming the perturbative
TFE approach for all disorder strengths. The method scales to system
sizes up to $L=48$, well beyond the reach of exact diagonalization,
demonstrating the practical utility of the tensor network formulation.

Beyond eigenenergies, VCUTs provides direct access to the diagonalizing unitary $U$ and the local
integrals of motion. The spatially resolved bond entropy $S^U$ of the diagonalizing unitary
reveals that, deep in the MBL phase, the entanglement
concentrates at bonds with small local disorder gradient $\Delta h_i$, identifying the
near-resonant bonds that govern the spatial structure of the emerging LIOMs. This provides a
direct real-space picture of the localization mechanism, consistent with the finite-depth circuit
structure expected for MBL systems~\cite{Pollmann2016}. The LIOMs obtained from VCUTs
accurately capture both short-time and long-time dynamics of the autocorrelation function,
and the eigenstate entanglement entropy statistics
provide insights into the ergodic-to-MBL crossover.

The main limitation of VCUTs lies in the ergodic regime, where the flow generates high entanglement
that exceeds the capacity of the finite bond dimension. This manifests as reduced accuracy at small
disorder and a mild shift of entanglement-based crossover diagnostics (such as the peak in
$\Delta_S^{\mathrm{samples}}$). Increasing the bond dimension systematically
improves the results, suggesting that the method can be refined to capture the transition region
more accurately.

Several directions for future improvement are apparent. First, the standard Wegner generator used
here becomes inefficient near degeneracies; incorporating scrambling
transformations~\cite{Thomson_2024} via a combination of Wegner and Toda-Mielke generators
(Appendix~\ref{section: Computing generators}) could enhance performance in such cases. Second,
advances in tensor network algorithms and high-performance computing would enable access to larger
bond dimensions and system sizes. Finally, the formalism developed here---based on vectorized MPOs
and TDVP evolution---is directly applicable to open quantum systems described by Lindblad master
equations~\cite{Zwolak_2004, Werner_2016, Guth_2020, Weimer_2021}, an extension we are currently
pursuing.

\begin{acknowledgements}
This work was funded by the Deutsche Forschungsgemeinschaft (DFG, German
Research Foundation) – 508440990. Simulations were performed with computing resources granted by
RWTH Aachen University under project rwth1807 and rwth1543.
\end{acknowledgements}
\begin{appendix}
\FloatBarrier
\begin{figure}
    \centering
    \includegraphics[width=1.\columnwidth]{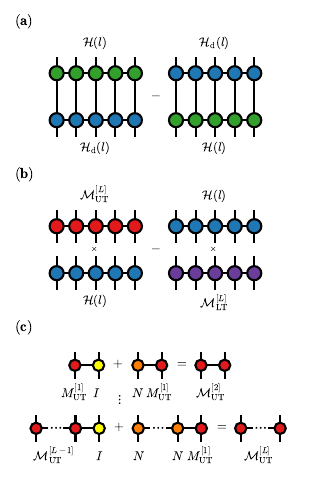}
\caption{\label{fig:demo for construct generators}
Tensor network construction of flow generators. (a)~Wegner generator
$\eta_{\mathrm{W}}(l) = [\ham_{\mathrm{d}}(l), \ham(l)]$ computed as the difference of two MPO
contractions. (b)~Toda-Mielke generator $\eta_{\mathrm{T}}(l)$ obtained via element-wise
multiplication (denoted $\times$) of the Hamiltonian with upper and lower triangular mask tensors.
(c)~Iterative construction of the upper triangular mask tensor $\mathcal{M}_{\mathrm{UT}}^{[L]}$
for a system of length $L$.}
\end{figure}
\section{Tensor network representations for generators}\label{section: Computing generators}
In this appendix we describe the tensor network construction of the Wegner and Toda-Mielke
generators used in continuous unitary transformations.

\subsection{Wegner generator}
The Wegner generator defined in Eq.~\eqref{eq:Wegner generator} can be equivalently written as
\begin{equation}
\label{Wegner generator II}
    \eta_{\mathrm{W}}(l) = \left[\ham_\mathrm{d}(l), \ham(l)\right],
\end{equation}
since $[\ham_\mathrm{d}(l), \ham_\mathrm{d}(l)] = 0$. This form is more convenient for
implementation because it requires only one commutator rather than explicitly separating
$\ham_{\mathrm{od}}(l)$.

The diagonal part $\ham_\mathrm{d}(l)$ is obtained by setting all off-diagonal elements within each
local $d \times d$ block of the physical indices to zero. The commutator is then computed using
standard MPO arithmetic, as illustrated in
Fig.~\ref{fig:demo for construct generators}(a). The resulting generator has bond dimension scaling
as $2D^2$, where $D$ is the bond dimension of the Hamiltonian MPO. To prevent memory growth during
the flow, we apply variational compression~\cite{Schollwoeck_2011, Jarno2011} to reduce the bond dimension, which
also avoids artificial memory effects that can arise from SVD-based truncation.

\subsection{Toda-Mielke generator}
The Toda-Mielke generator~\cite{Mielke_1998} is defined as
\begin{equation}
    \eta_{\mathrm{T}}^{\alpha\beta}(l) = \mathrm{sgn}(\beta-\alpha)\,\ham_{\alpha\beta}(l)\,,
    \label{def: Toda Generator}
\end{equation}
which preserves the band structure of the Hamiltonian while lifting degeneracies. Unlike the Wegner
generator, this requires constructing upper and lower triangular mask tensors in the many-body
Hilbert space. We use the $2 \times 2$ building blocks
$I = \begin{bmatrix} 1 & 0 \\ 0 & 1 \end{bmatrix}$,
$N = \begin{bmatrix} 1 & 1 \\ 1 & 1 \end{bmatrix}$, and
$M_{\mathrm{UT}} = \begin{bmatrix} 0 & 1 \\ 0 & 0 \end{bmatrix}$.
The upper triangular mask tensor $\mathcal{M}_{\mathrm{UT}}^{[L]}$ is constructed iteratively as
shown in Fig.~\ref{fig:demo for construct generators}(c), and the lower triangular mask
$\mathcal{M}_{\mathrm{LT}}^{[L]}$ follows analogously. Both masks are highly compressible to bond
dimension $D = 2$, independent of system size $L$. The Toda-Mielke generator is then obtained by
element-wise multiplication of $\ham(l)$ with these masks [Fig.~\ref{fig:demo for construct
generators}(b)]; details on element-wise MPO multiplication can be found in
Ref.~\cite{Shinaoka2023}.

\section{Tensor Flow Equations (TFE)}\label{sec: TFE}
TFE and VCUTs are both based on Wegner's flow equations but differ in their operator representation
and truncation scheme. In TFE, operators are expanded and truncated to a specified order in the
interaction strength $U$, yielding a resummed perturbative expansion that is most accurate at weak
interactions. For the TFE simulations presented in this work, we run the flow using a combination
of scrambling and Wegner generators until $l = 1000$ or until the convergence criteria
$\max|\ham^{(2)}_{\mathrm{off}}(l)| < 10^{-6}$ and $\max|\ham^{(4)}_{\mathrm{off}}(l)| < 10^{-4}$
are satisfied, where $\ham^{(2),(4)}_{\mathrm{off}}(l)$ denote the remaining off-diagonal elements
in the one- and two-particle sectors, respectively, as introduced in Ref.~\cite{Herre2024}. Since
TFE is formulated for fermionic operators, a Jordan-Wigner transformation is applied to express
Eq.~\eqref{eq:spinful_XXZmodel} in terms of interacting spinless fermions.
\end{appendix}

\FloatBarrier

\end{document}